\documentclass[final,3p,times,twocolumn]{elsarticle}

\usepackage{amsmath,amsthm,amsfonts,slashed,setspace}
\usepackage{epsfig,bbold,latexsym,yfonts,float}
\usepackage{color}
\usepackage[shortlabels]{enumitem}
\usepackage{gensymb,verbatim}
\usepackage{cancel}
\usepackage{multirow}
\usepackage{xcolor}
\usepackage{amssymb}
\usepackage{mathrsfs} %\mathscr{#}
\makeatletter
\def\mathcolor#1#{\@mathcolor{#1}}
\def\@mathcolor#1#2#3{%
  \protect\leavevmode
  \begingroup
    \color#1{#2}#3%
  \endgroup
}
\makeatother

\newcommand{\euv}{\varepsilon_{\rm UV}}

\usepackage{physics}

\newcommand{\nn}{\nonumber    \\ }

\journal{Physics Letters  B}

\begin{document}

%\begin{frontmatter}

\author{Ian Balitsky, Wayne Morris and  Anatoly  Radyushkin}
\address{Physics Department, Old Dominion University, Norfolk,
             VA 23529, USA}
\address{Thomas Jefferson National Accelerator Facility,
              Newport News, VA 23606, USA
}

\title{Gluon   Pseudo-Distributions  at Short Distances: Forward Case}

\begin{abstract}

%{Preprint JLAB-THY-19-}
%\bigskip

We  present the results that are  necessary in 
the ongoing  lattice calculations of the gluon parton distribution functions
(PDFs)  within  the pseudo-PDF approach.
We give a classification of possible two-gluon correlator functions and 
 identify  those  that  contain  the invariant amplitude 
  determining   the gluon PDF  in the light-cone $z^2 \to 0$ limit. 
One-loop calculations have been performed 
in the coordinate representation and 
in an explicitly 
  gauge-invariant form. 
We  made an effort to
separate  ultraviolet  (UV) and infrared (IR) sources of the  \mbox{$\ln (-z^2)$-dependence}
at short   distances $z^2$.  
The UV terms 
cancel in the   reduced  Ioffe-time distribution  (ITD),
and we obtain 
 the matching relation 
between the reduced ITD and the   light-cone  ITD. 
Using  a kernel form, we get  
a  direct connection between  lattice data for the  reduced ITD 
and the normalized 
gluon PDF.  We  also show  that our results 
may be used for a   rather   straightforward calculation  
of the 
one-loop matching relations for    quasi-PDFs.

%\vspace{5mm} 

%Keywords: Parton distribution  functions; Transverse  momentum; Quasi-distributions
              
\end{abstract}

%\pacs{11.10.-z,12.38.-t,13.60.Fz}
\maketitle

%\end{frontmatter}

\section{
 Introduction}

 Lattice calculations   of  parton distribution functions (PDFs)  are now a subject of considerable interest and efforts
(see Ref.  \cite{Cichy:2018mum}  for a recent review and references to extensive literature).
Modern  efforts  aim at the extractions of PDFs $f(x)$  themselves rather than 
their $x^N$ moments. 
On the lattice,  this  %%%%% is        possible   if one starts 
may be achieved by  switching from %%%%% 
local operators to current-current correlators  \cite{Detmold:2005gg}.   %%%%%  
A further idea is to start %%%%%%%%%%
with  equal-time correlators
%%%%%    the idea put forward  in 
\mbox{\cite{Braun:2007wv}  }. %%%%%%%%
%%% Refs.    \cite{Detmold:2005gg,Braun:2007wv} 
%%%%%   in application  to the current-current  correlators. 
 %%%%%  and  emphasized by 
 
 X. Ji,   in the paper  \cite{Ji:2013dva}  that strongly stimulated  further development,
 made a ground-breaking  proposal to consider equal-time versions of nonlocal operators %%%%%%%
 defining PDFs, distribution amplitudes, generalized  parton distributions, and %%%%%%%%%
 transverse momentum dependent distributions. %%%%%%%%%%
% Its 
In the case of usual PDFs, the %%%%%%%%%%%%
 basic concept  
 of Ji's approach %%%%%%%
 is  a    ``parton quasi-distribution''  (quasi-PDF) $Q(y,p_3)$  \cite{Ji:2013dva,Ji:2014gla},   
and  
%%%% usual 
PDFs are obtained from the  large-momentum $p_3 \to \infty$  limit 
of quasi-PDFs. 

Other approaches,   such as 
 the ``good lattice cross sections''
 \cite{Ma:2014jla,Ma:2017pxb}, the  Ioffe-time  analysis of equal-time correlators  \cite{Braun:2007wv,Bali:2017gfr,Bali:2018spj}
and  the pseudo-PDF approach \cite{Radyushkin:2017cyf,Radyushkin:2017sfi,Orginos:2017kos} 
are  coordinate-space oriented, and extract parton distributions  taking  the 
short-distance $z_3 \to 0$ limit.

Both the $p_3\to \infty$ and $z_3\to 0$ limits are  singular,
and one needs to use {\it matching relations} 
to 
 convert   the Euclidean  lattice data into
the usual   light-cone PDFs.  In the quasi-PDF approach, such relations were studied 
for quark    \mbox{\cite{Ji:2013dva,Xiong:2013bka,Ji:2015jwa,Izubuchi:2018srq}} and gluon 
 PDFs \cite{Wang:2017eel,Wang:2017qyg,Wang:2019tgg}, for  the
 pion distribution amplitude (DA)  \cite{Ji:2015qla} and generalized parton distributions (GPDs) 
\cite{Ji:2015qla,Xiong:2015nua,Liu:2019urm}.

Within the pseudo-PDF approach, 
the matching relations  were derived for  non-singlet 
PDFs   \cite{Ji:2017rah,Radyushkin:2017lvu,Radyushkin:2018cvn,Zhang:2018ggy,Izubuchi:2018srq}.
The strategy of the lattice extraction  of  non-singlet GPDs and the pion DA
using the pseudo-PDF methods was outlined 
in  a recent paper Ref.   \cite{Radyushkin:2019owq},
where the  matching conditions for these cases have been also derived. 
In the present paper, our main goal is to describe the basic points 
of the pseudo-PDF  approach to extraction of unpolarized gluon PDFs, and also  to find  
 one-loop  matching conditions. 

In the gluon case, the calculation is complicated by 
strict  requirements of gauge invariance. In this situation,
a very effective method is provided by the coordinate-representation approach of 
Ref. \cite{Balitsky:1987bk}. It is based on the background-field method 
and the  heat-kernel expansion. It allows, starting with the original gauge-invariant
bilocal operator, to find its  modification by one-loop corrections.
The results are obtained  in an  explicitly gauge-invariant form. 

In this approach,   there is  no need 
to specify the  nature of  matrix element characteristic of   a particular 
 parton distribution.  This means that  one and  the same Feynman diagram 
 calculation may be used both for finding 
 matching conditions for PDFs (given by forward matrix elements), 
 and for  DA's and GPDs  corresponding to non-forward ones  (see Ref.  \cite{Radyushkin:2019owq}
 for an illustration of how this works  for  quark operators).

  The paper is organized as follows. 
In 
 \mbox{Section  2},   we analyze  the kinematic structure  of the matrix  elements of the gluonic bilocal operators,
 and identify those that contain information about the twist-2 gluon PDF.
 
Next,  we discuss one-loop corrections.  In \mbox{Section 3,}  we analyze  the  gauge-link self-energy contribution 
 and specific properties of its ultraviolet  and short-distance behavior.
 Our results for the vertex corrections to the gluon link are given  in \mbox{Section 4} 
 in the form that is valid both in forward and non-forward cases. 
 The ``box'' diagram is discussed in Section 5. 
 Since our results in this case are rather lengthy,
 we present just  some of them, and  in the forward case only. 
 The gluon self-energy corrections are discussed in Section 6. 
 
 The subject of Section 7 is the structure  of  perturbative evolution 
 of the gluon operators and matching conditions.
 Section 8     contains a   summary of the paper.

\setcounter{equation}{0}

%\newpage

\section{Matrix elements}

The nucleon spin-averaged  matrix elements for operators composed of  two-gluon-fields   (with all
 four indices  non-contracted)   are specified  by 
    \begin{align}
 { M}_{\mu \alpha;  \lambda \beta }  (z,p) \equiv \langle  p |  \,
 G_{\mu \alpha} (z)  \, [z,0]\,  G_{ \lambda \beta } (0) | p \rangle \  , 
\end{align}
 where  $
[z,0]$ is  the  standard  straight-line gauge link 
 in the gluon  (adjoint) 
 representation
 \begin{align}
[x,y]~\equiv~{\rm Pexp}\Big\{ig\!\int_0^1\! dt~(x-y)^\mu \tilde{A}_\mu(tx+(1-t)y)\Big\}
 \  . 
 \label{straightE}
\end{align}
 The tensor structures for   a decomposition over   invariant amplitudes    
 may be built from two  available 4-vectors $p_\alpha$, $z_\alpha$  and the 
 metric tensor $g_{\alpha \beta}$. Incorporating the antisymmetry   of $G_{\rho \sigma}$ with  respect to its indices, we have 
    \begin{align}
& { M}_{\mu \alpha;   \lambda  \beta}  (z,p)= \nonumber \\  & 
 \left( g_{\mu\lambda} p_\alpha p_\beta - g_{\mu\beta} p_\alpha p_\lambda - g_{\alpha\lambda} p_\mu p_\beta + g_{\alpha\beta} p_\mu p_\lambda \right) \mathcal{M}_{pp}  \nonumber \\
+& \left( g_{\mu\lambda} z_\alpha z_\beta - g_{\mu\beta} z_\alpha z_\lambda - g_{\alpha\lambda} z_\mu z_\beta + g_{\alpha\beta} z_\mu z_\lambda \right) \mathcal{M}_{zz}     \nonumber \\
+ & \left( g_{\mu\lambda} z_\alpha p_\beta - g_{\mu\beta} z_\alpha p_\lambda - g_{\alpha\lambda} z_\mu p_\beta + g_{\alpha\beta} z_\mu p_\lambda \right) \mathcal{M}_{zp}    \nonumber \\
+ & \left( g_{\mu\lambda} p_\alpha z_\beta - g_{\mu\beta} p_\alpha z_\lambda - g_{\alpha\lambda} p_\mu z_\beta + g_{\alpha\beta} p_\mu z_\lambda \right) \mathcal{M}_{pz}     \nonumber \\
+&   \left( p_\mu z_\alpha  - p_\alpha z_\mu\right) \left(  p_\lambda z_\beta -  p_\beta z_\lambda\right) \mathcal{M}_{ppzz} 
\nn  + &   \left(g_{\mu\lambda} g_{\alpha\beta} -g_{\mu\beta} g_{\alpha\lambda} \right)\mathcal{M}_{gg}    \ ,
\label{Manb}
\end{align}
where the  amplitudes ${\cal M}$ are functions of 
the invariant interval $z^2$ and the Ioffe time \cite{Ioffe:1969kf} 
$(pz)\equiv - \nu$   (the minus sign
is introduced for further  convenience).

Since the matrix element should be symmetric with respect to interchange of the
fields (which amounts to $\{\mu \alpha\} \leftrightarrow \{ \lambda \beta \}$ and $z \to -z$), 
the  functions $\mathcal{M}_{pp}$,  $\mathcal{M}_{zz}$, $\mathcal{M}_{gg} $,  $\mathcal{M}_{ppzz} $ and 
$ \mathcal{M}_{pz} - \mathcal{M}_{zp}  $  are
even functions of $\nu$, while   $ \mathcal{M}_{pz} + \mathcal{M}_{zp}  $ is
odd in $\nu$.

The usual light-cone gluon distribution is obtained from 
 $ g^{\alpha \beta} { M}_{+ \alpha;  \beta  +}  (z,p) $,  with $z$ taken in the light-cone ``minus'' direction,
 $z=z_-$.  We have
  \begin{align}
 g^{\alpha \beta} { M}_{+ \alpha;  \beta +}  (z_-,p) =-2 p_+ ^2  \mathcal{M}_{pp}(\nu ,0)  \ , 
 \label{lcpp}
\end{align}
i.e.,  the  PDF  is  determined by the ${\cal M}_{pp} $ structure,
  \begin{align}
- \mathcal{M}_{pp} (\nu,0) =\frac12
 \int_{-1}^1 \dd x \, e^{-i x \nu} x f_g (x) \,  \ . 
 \label{glPDF}
\end{align}
Thus, we should choose the operators with the sets
$\{\mu \alpha; \lambda \beta  \}$  that contain ${\cal M}_{pp} $
in their parametrization. 

Note that  it is  the density of the momentum \mbox{$G(x) \equiv x f_g (x)$}  
 carried by the gluons rather than their number
density  $f_g (x)$ that is a natural quantity in this  definition of  the gluon  PDF.
In the local $z_-=0$  (or $\nu=0$) limit, the  $x$-integral gives the fraction of the 
hadron's plus momentum carried by the gluons.
In the absence of gluon-quark transitions, this fraction is conserved,
which puts a restriction on the $gg$-component of the Altarelli-Parisi \cite{Altarelli:1977zs} kernel.
Namely, it should have the plus-prescription 
property when applied to $G(x)$.

Due to  antisymmetry of $G_{\rho \sigma}$ with  respect to its indices,
the values $\alpha=+$ and $\beta= +$ are excluded from the summation in Eq. (\ref{lcpp}). 
Furthermore, since $g_{- -} =0$, the combination  
$g^{\alpha \beta}  { M}_{+ \alpha; \beta +}  (z,p)$ includes only summation over 
transverse indices $i,j =1,2$, i.e. reduces 
to $g^{ij}  { M}_{+ i;  j +}  (z,p) \equiv  { M}_{+ i;  + i}  (z,p)  $ 
(we switched here to Euclidean summation over $i$), for which we have 
  \begin{align}
 { M}_{+ i;  +i}  
 =   { M}_{0 i;  0i}  +    { M}_{3 i;   3i}  +   ({ M}_{0 i;  3i}  +
 { M}_{3 i;    0i}  ) \ . 
\label{ii}
\end{align}
In the local  $z_3=0$  limit, these three combinations are proportional to ${\bf E}^2_\perp$, ${\bf B}^2_\perp$
and the third component $({\bf E \times B})_3$ of the Poynting vector,
respectively.

The decomposition of these combinations (with summation over $i$) in the basis of   the  ${\cal M}$ structures is 
\begin{align}
 { M}_{0 i;  i 0 }   =  & 2   p_0^2  \mathcal{M}_{pp}   +2 \mathcal{M}_{gg}  \ ,  \\ 
   { M}_{3 i;  i 3 }   =  &2   p_3^2  \mathcal{M}_{pp}+2   z_3^2  \mathcal{M}_{zz}   \nonumber \\ &
 +2  z_3 p_3  \left( \mathcal{M}_{zp} +   \mathcal{M}_{pz} \right) - 2\mathcal{M}_{gg}  \ , 
\\ 
{ M}_{0 i;   i 3 }  = & 2 p_0 \left(    p_3  \mathcal{M}_{pp}  + z_3  \mathcal{M}_{pz}  \right)  \ , 
\\ 
 { M}_{3 i;  i 0 }   =  & 2 p_0 \left(     p_3  \mathcal{M}_{pp}  +  z_3   \mathcal{M}_{zp}  \right)  \  . 
\end{align}

All of them contain the $ \mathcal{M}_{pp} $ function defining the gluon 
distribution, though with different kinematical factors.
Unfortunately, none of them is just $ \mathcal{M}_{pp} $:  they all contain 
contaminating  terms. 
Moreover,  the $   { M}_{3 i; i 3} $ matrix element (proposed 
originally  \cite{Ji:2013dva}  for extractions of the gluon PDF on the lattice) 
contains three contaminations, while the  others have 
just one  addition.
 In particular, the matrix element ${ M}_{0 i;  i 0  } $  has   $ \mathcal{M}_{gg}  $ as a contaminating term.
It is easy to see that 
\begin{align}
 { M}_{  j i ; ij  }  \equiv  \bra{p} G_{j i } (z) G_{ij}(0) \ket{p} &= -2  \mathcal{M}_{gg}  \ , 
\end{align}
where the   summation over both $i$ and $j$ is assumed. Hence, the combination 
\begin{align}
 { M}_{0 i;  i 0 }  +  { M}_{j i ;  i j } =  & 2   p_0^2  \mathcal{M}_{pp}   \ 
 \label{00m}
\end{align}
may be used for extraction
of the twist-2 function $ \mathcal{M}_{pp} $.

Combining together matrix elements of different types, one should 
 take into account 
that,  
off the light cone, these matrix elements have extra  ultraviolet divergences
related to presence of the gauge link.
Due to   the local nature of   ultraviolet divergences,
each matrix element, for any   set of its indices $\{ \mu \alpha;  \lambda \beta  \}$, 
 is multiplicatively renormalizable with respect to these divergences \cite{Li:2018tpe}.
However,  choosing different sets of  $\{\mu \alpha; \nu \beta\}$, we get, in general,  different 
 anomalous dimensions. 
 
Thus, it is not evident {\it a priori} which  linear combinations of these 
matrix elements are  multiplicatively renormalizable. 
In Ref. \cite{Zhang:2018diq}, it was established that  the combinations  represented in Eq. (\ref{ii}),  namely,  
$ { M}_{0 i; i 0 }  $,    $ { M}_{3 i;  i 3 } $,    \mbox{${ M}_{0 i; i3 }  +
 { M}_{3 i; i 0 }  $} (and also  \mbox{${ M}_{0 i; i 3}  -
 { M}_{3 i; i 0 }  $}),  with   
 summation over     transverse indices $i$,  
 are each  multiplicatively renormalizable at the one-loop level. 
 
 Furthermore,  the combination $G_{ij}G_{ij}$  (with summation over  transverse $i,j$) 
  equals to $2G_{12} G_{12}$,
 whose matrix elements are  multiplicatively renormalizable.
 As we will see, it has the same one-loop UV anomalous dimension
 as  ${M}_{0 i;  i 0 } $, hence   the combination of  Eq. (\ref{00m}) 
 is multiplicatively renormalizable at the one-loop level. 
 A possible subject for further studies is to investigate %%%%%%%
 if this is true in higher orders. %%%%%%%%%%

 The combination $g^{\alpha \beta}  { M}_{3 \alpha; 3 \beta } $, 
 containing a covariant summation  over $\alpha$ and $\beta$, 
 was also  found to be multiplicatively renormalizable. 
 It is given by 
\begin{align}
& g^{\alpha \beta}   { M}_{3 \alpha; 3 \beta } =  
 \left(2p_3^2 - m^2    \right) \mathcal{M}_{pp} + 3z_3^2  \mathcal{M}_{zz}   
 \nonumber \\ & +  3 p_3 z_3  \left( \mathcal{M}_{zp}+  \mathcal{M}_{pz}    \right) +   p_0^2 z_3^2  \mathcal{M}_{ppzz}
-3\mathcal{M}_{gg}  \ , 
\end{align}
and has the largest number (four) of    contaminations. 

The function $g^{\alpha \beta}  { M}_{0 \alpha; 0 \beta } $, also  
involving a covariant summation,  was used in the first attempt
\cite{Fan:2018dxu}
of the lattice extraction of the gluon PDF. However, as noted in Ref.  
\cite{Zhang:2018diq}, it is not multiplicatively renormalizable.

In any  theory with a dimensionless coupling constant,  
the matrix elements $ { M} (z,p)$ 
 contain  \mbox{$\sim \ln (-z^2)$}  terms 
corresponding to      perturbative (or ``DGLAP'' for Dokshitzer-Gribov-Lipatov-Altarelli-Parisi 
\cite{Gribov:1972ri,Altarelli:1977zs,Dokshitzer:1977sg}) 
evolution. One may wonder which  combinations  have   
a diagonal DGLAP evolution  at one loop. 

To answer these questions, we have 
calculated the modification of the original bilocal operator by 
one-loop 
 gluon exchanges.

 \setcounter{equation}{0}

\section{Link self-energy  contribution and ultraviolet divergences}

\begin{figure}[t]
   \centerline{\includegraphics[width=2in]{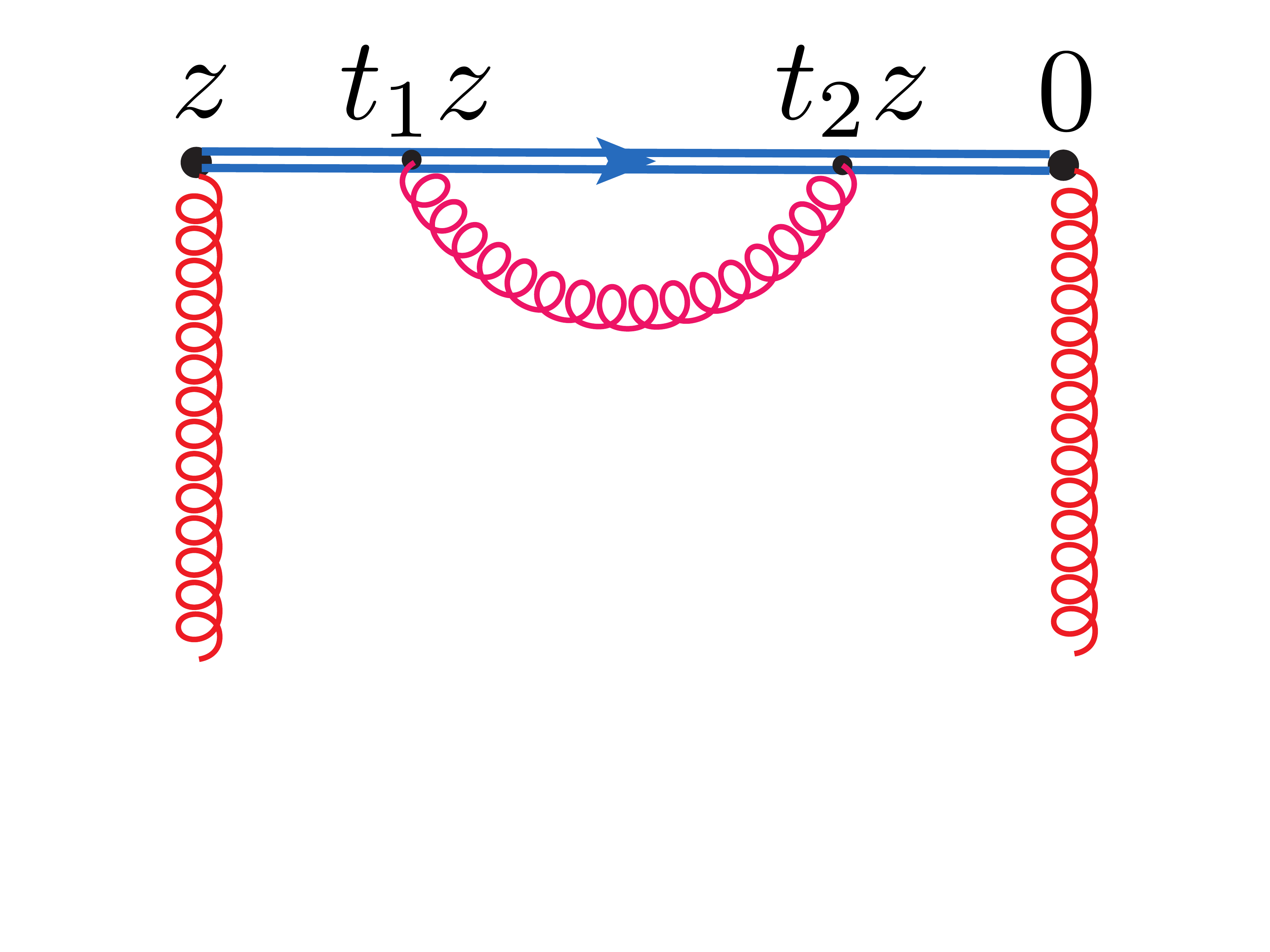}}
        \vspace{-8mm}
   \caption{Self-energy-type  correction for  the gauge link.
   \label{linkself}}
   \end{figure}

The simplest diagram corresponds to  the 
self-energy correction  for  the  gauge 
 link (see Fig. \ref{linkself}).  Its  calculation is the same as in  case of the quark bilocal operators
 (see, e.g., Ref. \cite{Radyushkin:2017lvu}).
 At one  loop,  one should just  the change the color factor $C_F \to C_A$. Thus, we have 
  \begin{align} 
 \Gamma_ \Sigma (z)  = &  ({i} g)^2\,C_A \,\frac12 \,  \int_0^1 \dd t_1 \,  \int_0^1 \dd t_2 \, 
  z^\mu z^\nu \,  D_{\mu \nu} ^c [ z (t_2-t_1) ] \  , 
    \label{self}
 \end{align}
  where    $D_{\mu \nu} ^c (z)  =g_{\mu \nu}  /4\pi^2 z^2$ 
  is the Feynman-gauge gluon propagator in the coordinate representation.
The resulting  integrals over the link parameters $t_1, t_2$      \begin{align} 
     \int_0^1 \dd t_1 \,  \int_0^1 \frac{\dd t_2}{(t_2-t_1)^2}  \, 
    \label{t1t2}
 \end{align}
 diverge 
when   $t_1 \sim t_2$, i.e.,  when 
 the endpoints $t_1z$ and  $t_2 z$ of  the gluon propagator  are close to each other. So,  
 one may suspect that this divergence has an ultraviolet origin. 
  To  see that this is the case, we  use the dimensional regularization  (DR) \cite{tHooft:1972tcz}
in the UV region, switching to
$d$ dimensions.
As a result,  the gluon propagator in  the coordinate space  acquires  an extra factor  $(-z^2)^{ 2-d/2}$.
This  results in an extra $(t_2-t_1)^{2-d/2}$ factor in Eq. (\ref{t1t2}), and the integral there converges
for sufficiently small $d$.

 To preserve gauge invariance, our calculations were made using 
 massless gluons  and the dimensional regularization.
 However,     in   the case of the link self-energy diagram, the use of DR
 (which  is  basically just a  mathematical trick)  is rather misleading in a couple of points.
 
 The relevant subtleties may be   illustrated by using the Polyakov prescription 
 \mbox{$1/z^2 \to 1/(z^2-a^2)$}  for the gluon propagator in the coordinate representation   \cite{Polyakov:1980ca} 
    (see also Refs.  \cite{Chen:2016fxx,Radyushkin:2017lvu}).
 It 
softens the gluon propagator at  intervals  $-z^2 \lesssim a^2 $,  
and eliminates its singularity at $z^2=0$. In this respect,
it is similar to  the UV regularization  produced by a  finite lattice spacing, and gives   
  \begin{align} 
  \Gamma_ \Sigma (z,a)  = & - g^2\,C_A \,\frac{z^2}{8 \pi^2}
   \,  \int_0^1 \dd t_1 \,  \int_0^1  \, \frac{\dd t_2}{ z^2 (t_2-t_1)^2 - a^2}  \  . 
    \label{selfa}
 \end{align}
 The regularized integral   vanishes 
 on the light cone \mbox{$z^2=0$}  and converges for spacelike $z$.  
Taking $z=z_3$ and 
  calculating the  integrals gives  \cite{Chen:2016fxx,Radyushkin:2017lvu}
  \begin{align} 
 \Gamma_  \Sigma (z_3,a)  =   -\,C_A \,\frac{\alpha_s}{2 \pi}
  & 
  \left [ 
   \,
 2  \frac{ z_3}{a} \,  \tan
   ^{-1}\left(\frac{z_3}{a}\right)   -  \ln 
   \left(1+ \frac{z_3^2}{a^2}\right) \right ] \   .
       \label{selfex}
        \end{align}
   The result contains a linear $ \sim 1/a$ divergence  that is 
missed  if one uses the  DR. Furthermore, 
for a {\it  fixed}  $a$ and small $z_3$ it  behaves 
 like $z_3^2/a^2$, i.e.,  $  \Gamma_  \Sigma (z,a)  $  vanishes for $z_3=0$,  as expected:  there is no link if $z_3=0$. 
  It also vanishes on the light cone $z^2=0$.     
  
 The fact that     $ \Gamma_  \Sigma (z_3=0,a) =0$  means that, 
 for  a  fixed $a$,  %%%%%%%%
 this term   gives  no corrections to
the local limit of the $ G_{\mu \alpha} (z)  \, [z,0]\,  G_{ \lambda \beta } (0)$   %%%%%
operator, e.g., to %%%%%%%%
 the energy-momentum tensor (EMT).  
 %% for  a  fixed $a$.  
 Since the matrix element 
 of the EMT gives  the  fraction of the hadron momentum carried by the gluons, 
 the link self-energy correction  does not change this fraction. This  is a  natural 
 phenomenon in the absence of 
 the gluon-quark transitions.

 However, if one formally takes the $a\to 0$ limit for a fixed $z_3$ in Eq. (\ref{selfex}),
 then $ \ln    \left(1+{z_3^2}/{a^2}\right)$  converts into the expression 
  $ \ln    \left({z_3^2}/{a^2}\right)$  singular for $z_3=0$. 
    Similarly, using the DR,  one faces an  outcome proportional to   
  \begin{align} 
   (-z^2\mu_{\rm UV}^2)^{\euv}/\euv= 1/\euv + \ln (-z^2 \mu_{\rm UV}^2) + \ldots \, , 
     \end{align} 
where $\mu_{\rm UV}$ is the scale accompanying this UV dimensional regularization.  
Again, the starting expression vanishes for $z^2=0$, but %%%%%%%% 
%%%   Thus, 
 renormalizing it by  a  subtraction of the $1/\euv$ pole, %%%%%%%%
one may apparently  conclude that, in addition to the UV divergence, this diagram 
 contains a singularity on the light cone $z^2=0$.
 
   For this reason, in our DR results  we will explicitly separate 
the \mbox{$z^2$-dependence}  induced by the UV singular terms 
(that actually  vanish on the light cone) %%%%%%%%
 %%%%%  if one uses a more physical regularization, such as lattice spacing), 
and
 that present in the DGLAP-evolution 
logarithms $\ln (-z^2 \mu_{\rm IR}^2)  $, where $ \mu_{\rm IR}$ is the  scale 
associated with  the DR regularization of the collinear singularities. 

The main difference is  that if, instead of DR, one regularizes collinear singularities
by using a physical IR cut-off $\Lambda$ (like nonzero gluon virtuality or gluon mass),
the one-loop result,    proportional to the modified Bessel function $K_0 (\sqrt{-z^2 \Lambda^2})$,    remains 
 singular for $z^2=0$, unlike the UV-induced logarithm
$\ln (1-z^2/a^2)$.

 In  the case of  the link self-energy diagram, we have  UV singularities only.
Its correction to the $G_{\mu \alpha}(z)G_{\lambda \beta }(0)$ operator is given by 
 \begin{align}
&
-{g^2N_c \over 4\pi^2[(-z^2 \mu_{\rm UV}^2 +i\epsilon)]^{{d\over 2}-2}} {\Gamma\big({d/ 2}-1\big) \over (3-d)(4-d)}G_{\mu\alpha}(z)G_{\lambda \beta }(0) \ ,
\label{selfAD}
\end{align}
where the $1/(3-d)(4-d)$ factor  results 
from the  integral 
 \begin{align}
&
   \!\int_0^1\! \dd t_1 \!\int_0^{t_1} \! \dd t_2\,  (t_1-t_2)^{2-d} ={1 \over (3-d)(4-d)}  
\nonumber
\end{align}
produced by the DR of the  gluon propagator \mbox{$D^c (t_1z - t_2 z)$.}
The pole for $d=3$ ($d=4$) corresponds to the linear (logarithmic)  UV divergence
in Eq. (\ref{selfex}).

\setcounter{equation}{0}

\section{Vertex contributions}

There are also   vertex diagrams     
involving gluons  that connect the gauge link with the gluon  lines, see  \mbox{Fig. \ref{link}. } 

We  use  the method of calculation %%%%%
described in Ref.  \cite{Balitsky:1987bk}. It is  based on the background-field technique, %%%%%
with the gluon propagator taken %%%%%
in the ``background-Feynman'' (bF) gauge \cite{Balitsky:1987bk}.  %%%%%
It should be noted that the three-gluon vertex in the bF gauge is different %%%%%
from the usual Yang-Mills vertex (see e.g. \cite{Abbott:1980hw}). %%%%%
Therefore, the results obtained for separate diagrams in the %%%%%
bF gauge  differ from  those obtained in  the usual Feynman gauge %%%%%
and only the sum of all diagrams must be the same. %%%%%

   \subsection{UV divergent term} 

\begin{figure}[t]
   \centerline{\includegraphics[width=2.5in]{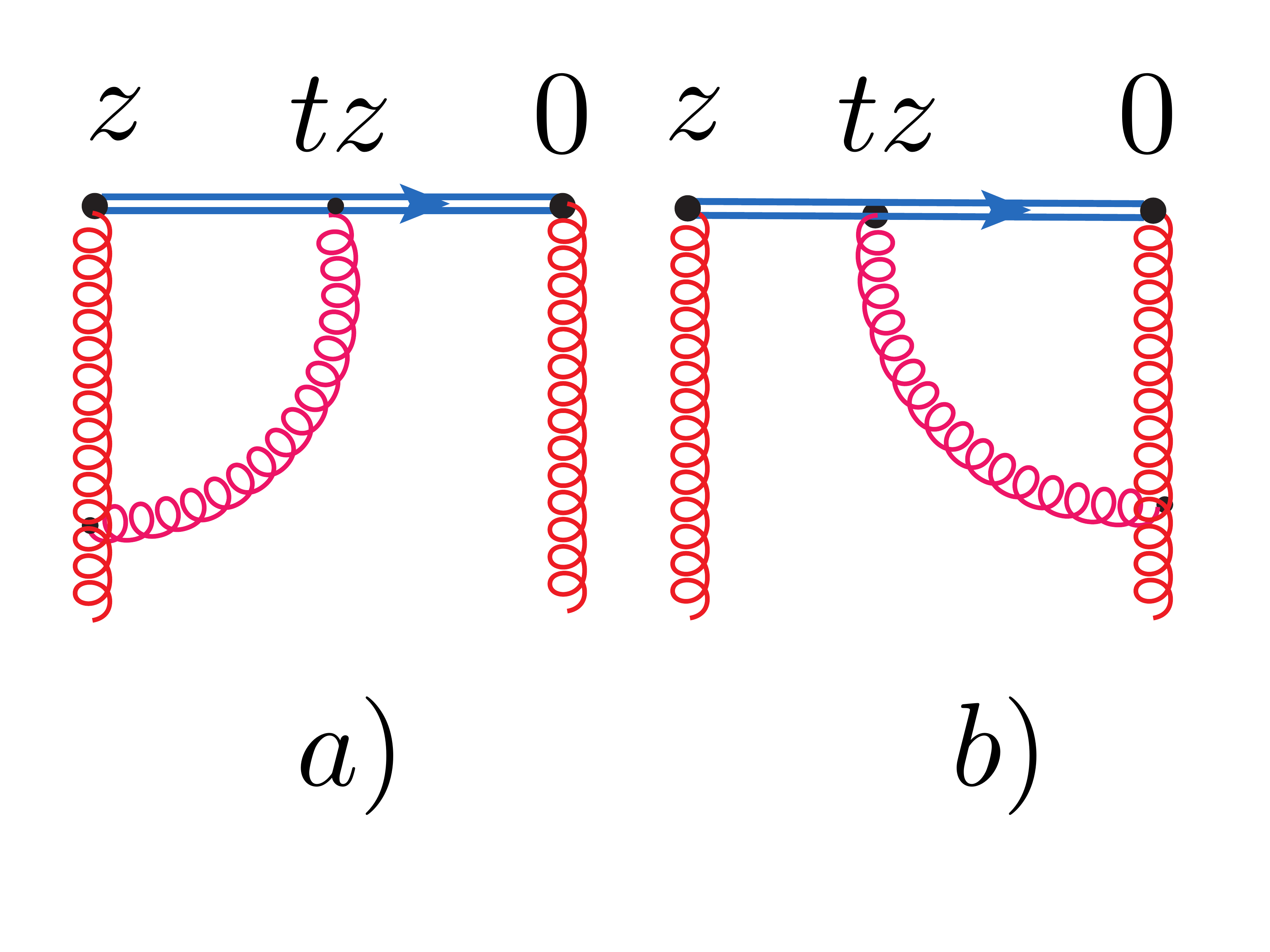}}
        \vspace{-5mm}
   \caption{Vertex diagrams with  gluons coming out of the gauge link.
   \label{link}}
   \end{figure}

Clearly, the gluon exchange produces a correction just to one of the fields in the 
$G_{\mu\alpha}(z)G_{\lambda \beta }(0)$ operator, while another remains intact. 
In particular, the diagram \ref{link}a changes $G_{\mu\alpha}(z)$ into the sum of two
terms.  One of them contains UV divergences, while the other one is UV finite. 

The UV-divergent  term is   given by 
\begin{align}
&\frac{N_c g^2}{4\pi^2} 
\frac{\Gamma(d/2-1)}{(d-2 )(-z^2)^{d/2-1}}
\int_0^1 \dd u\,  \left(u^{3-d}-u\right)  \nn  &\quad  \times 
 \left(z_{\alpha} G_{z\mu}( \bar uz) - z_{\mu} G_{z\alpha}(\bar uz)\right) \ , 
 \label{2a1}
 \end{align}
 where $G_{z\sigma} \equiv z^\rho G_{\rho\sigma}$ and $\bar u \equiv 1-u$.  
 The overall
 \mbox{$d$-dependent}  factor  here is finite for $d=4$,  but  the \mbox{$u$-integral} diverges at the lower limit.
 Thus, just like in the case of the link self-energy diagram, the divergence 
appears in the integral over a dimensionless parameter $t$ specifying the location 
of the endpoint of the gluon line on the gauge link.
The divergence  disappears if one uses the UV    regularization by  taking 
 $d=4-2\varepsilon_{\rm UV}$,  which converts it 
 into a pole at $\euv=0$.

 Since the UV  divergence comes from the \mbox{$u \to 0 $}  integration, we 
 can isolate it by taking $\bar u =1$ in the gluonic field, which gives 
 \begin{align}
&\frac{N_c g^2}{4\pi^2} 
\frac{\Gamma(d/2-1)}{(d-2 )(-z^2)^{d/2-1}}
\left (\frac1{4-d}-\frac12 \right )  \nn &\quad  \times 
 \left(z_{\alpha} G_{z\mu}(z) - z_{\mu} G_{z\alpha}(z)\right) \ . 
 \label{UVvert}
 \end{align}
 The remainder is given by 
 \begin{align}
&\frac{N_c g^2}{4\pi^2} 
\frac{\Gamma(d/2-1)}{(d-2 )(-z^2)^{d/2-1}}
\int_0^1 \dd u\,  \left[u^{3-d}-u\right]_{+(0)}  \nn  &\quad  \times 
 \left(z_{\alpha} G_{z\mu}( \bar uz) - z_{\mu} G_{z\alpha}(\bar uz)\right) \ , 
  \label{UVvertreg}
 \end{align}
 where the plus-prescription at $u=0$  is defined as
   \begin{align}
& \int_0^1  \dd u \left[f(u)\right]_{+(0)} g(u) = \int_0^1  \dd u f(u) [g(u) -g(0)]  \ .
 \label{plus0}
 \end{align}

At first sight,  the field ${\cal G}_{\mu\alpha}(z)=
z_{\alpha} G_{z\mu}( z) - z_{\mu} G_{z\alpha}(z)$
accompanying the UV pole in Eq. (\ref{UVvert}) does not look like
the field 
$ G_{\mu\alpha}(z)$ in the original operator. 
Thus, one may worry that we are not dealing here with a 
multiplicative UV renormalization.
 So, let us perform an explicit check for our particular case when 
 $z= \{0,0,0,z_3 \}$.   
 
 To begin with,  we see that  ${\cal G}_{\mu\alpha}(z)=0$  when  both $\mu$ and $\alpha$ 
are transverse indices $i,j$. This corresponds to a multiplicative renormalization
with the anomalous dimension (AD) equal to zero. 

Take now 
 $\mu=0$.  Then  ${\cal G}_{0\alpha}(z)= z_{\alpha} G_{z 0}( z) $, so that 
${\cal G}_{0 i}(z)= 0$ while ${\cal G}_{0 3}(z)= -z_3^2 G_{3 0}( z) = z_3^2 G_{0 3}( z)$. 
Finally, if $\mu=3$, then  ${\cal G}_{3\alpha}(z)= - z_{3} G_{z \alpha}( z) = z_3^2 G_{ 3 \alpha }( z) $,
which gives ${\cal G}_{3 i}(z)=  z_3^2 G_{ 3 i }( z) $ and  ${\cal G}_{3 0}(z)= z_3^2 G_{3 0}( z) $
(same  result as above). 

Thus, for all the cases, ${\cal G}_{\mu\alpha}(z)$ is a multiple of 
 ${G}_{\mu\alpha}(z)$.  Namely, when one of the 
indices equals 3, 
 we have a nontrivial anomalous dimension, since 
${\cal G}_{3\alpha}(z)=- {\cal G}_{\alpha 3}(z)=   z_3^2 G_{ 3 \alpha }( z) $.
In all other cases, we have a trivial (vanishing)  AD, since   ${\cal G}_{ij}(z)=0$ and  ${\cal G}_{0i}(z)=0$. 

As mentioned, the link self-energy diagram has both  linear and logarithmic 
UV divergences, while the vertex diagrams have  just  logarithmic 
UV divergences. 
Adding the logarithmic UV divergence  coming from the link self-energy
to the   UV divergences of the vertex diagrams, we find, in particular,  that 
the  matrix elements $ { M}_{0 i; i 0 }  $ and $ { M}_{ i j; i j}  $ have  the logarithmic  AD   due to the link self-energy diagram only.
Call it $ \gamma$.  
 Comparing overall factors in  Eqs. (\ref{selfAD}) and  (\ref{UVvert}), we conclude that 
   $ { M}_{3 i; i 3 } $  has   the  logarithmic AD   equal to $2 \gamma$ and  matrix elements
      \mbox{${ M}_{0 i; i 3 }  \pm 
 { M}_{3 i;  i 0 }  $}  have  the   logarithmic AD  equal to $\frac32  \gamma$.
 In addition, all of these structures acquire at one loop  the same factor due to the linear UV singularity.

   \subsection{Evolution term}
 
Our   calculations show that   the second, UV finite term from  the diagram \ref{link}a  is   given by 
 \begin{align}
\quad   \frac{N_c g^2}{8\pi^2} \frac{\Gamma(d/2-2)}{({d-3})(-z^2)^{d/2-2}} &
 \int_0^1 \dd u   \left[u^{3-d}-1\right]_{+(0)}  \nonumber 
 \\ &\quad  \times  
  G_{\mu\alpha}( \bar uz)  G_{\lambda \beta }( 0) \ .
\label{2aEv}
\end{align}
Note that the gluonic operator in Eq. (\ref{2aEv}) 
has the same tensor structure as the original operator 
$G_{\mu\alpha}( z)  G_{\beta \nu}( 0)$ differing from it just by 
rescaling $z \to \bar u z$. 
 There is no mixing with operators  of a different type. 
 The $u$-integral in this case does not diverge   for $d=4$, but the  overall 
\mbox{$\Gamma(d/2-2)$}  factor  has a pole  $1/(d-4)$.

Formally,  there is also a pole   $1/(d-3)$,   %%%%%%%
corresponding to a linear UV divergence. %%%%%%%%%
However,  the singularity %%%%%%%%%
for $d=3$ is eliminated by the $\left [u^{3-d}-1\right ]$ combination in the integrand. %%%%%%%%%
One may say that the linear divergences present in ``$u^{3-d}$''  and ``$-1$''  %%%%
parts cancel each other.  %%%%

In the calculation of Refs.\cite{Zhang:2018diq,Wang:2019tgg} performed using  the usual Feynman gauge, %%%%
the linear singularities cancel between contributions %%%%
of two different diagrams shown in Fig. 1 of Ref. \cite{Wang:2019tgg}.   %%%%
In our calculation,  based on the bF gauge, 
%(see Appendix A),
 the sum of these   diagrams is  represented by  just one vertex diagram, %%%%
so that the cancellation %%%%
occurs inside the contribution (\ref{2aEv}) of   that  diagram. %%%%

The remaining  $1/(d-4)$  pole  %%%%
%% It 
corresponds to a collinear divergence 
developed %%%%%
 because  all  the propagators correspond to massless particles.
Taking a nonzero gluon mass $\lambda$, one would get a  finite  result 
containing $K_0(\sqrt{-z^2} \lambda)$
(see, e.g., Ref. \cite{Radyushkin:2017lvu} for a discussion of 
the quark vertex diagram in a similar context). 

Still, $K_0(\sqrt{-z^2} \lambda)$  is only finite as far as $z^2$ is finite.
The  IR cut-off  does not eliminate
the logarithmic singularity $\ln (-z^2 \lambda^2)$ that 
$K_0(\sqrt{-z^2} \lambda)$ has on the light cone. 
In  the $z=z_3$ case, 
 $z_3^2$  works like an  ultraviolet cut-off
 for this singularity. 
 This may be  contrasted with   the UV divergent contributions, 
 where the UV cut-off is provided by the Polyakov regularization parameter $a$
 (or lattice spacing $a_L$)  while $z_3^2$ appears on the IR side 
 of the relevant logarithm $\ln (z_3^2/a^2)$.

\setcounter{equation}{0}

\section{Box diagram}

  \begin{figure}[b]
   \centerline{\includegraphics[width=1.7in]{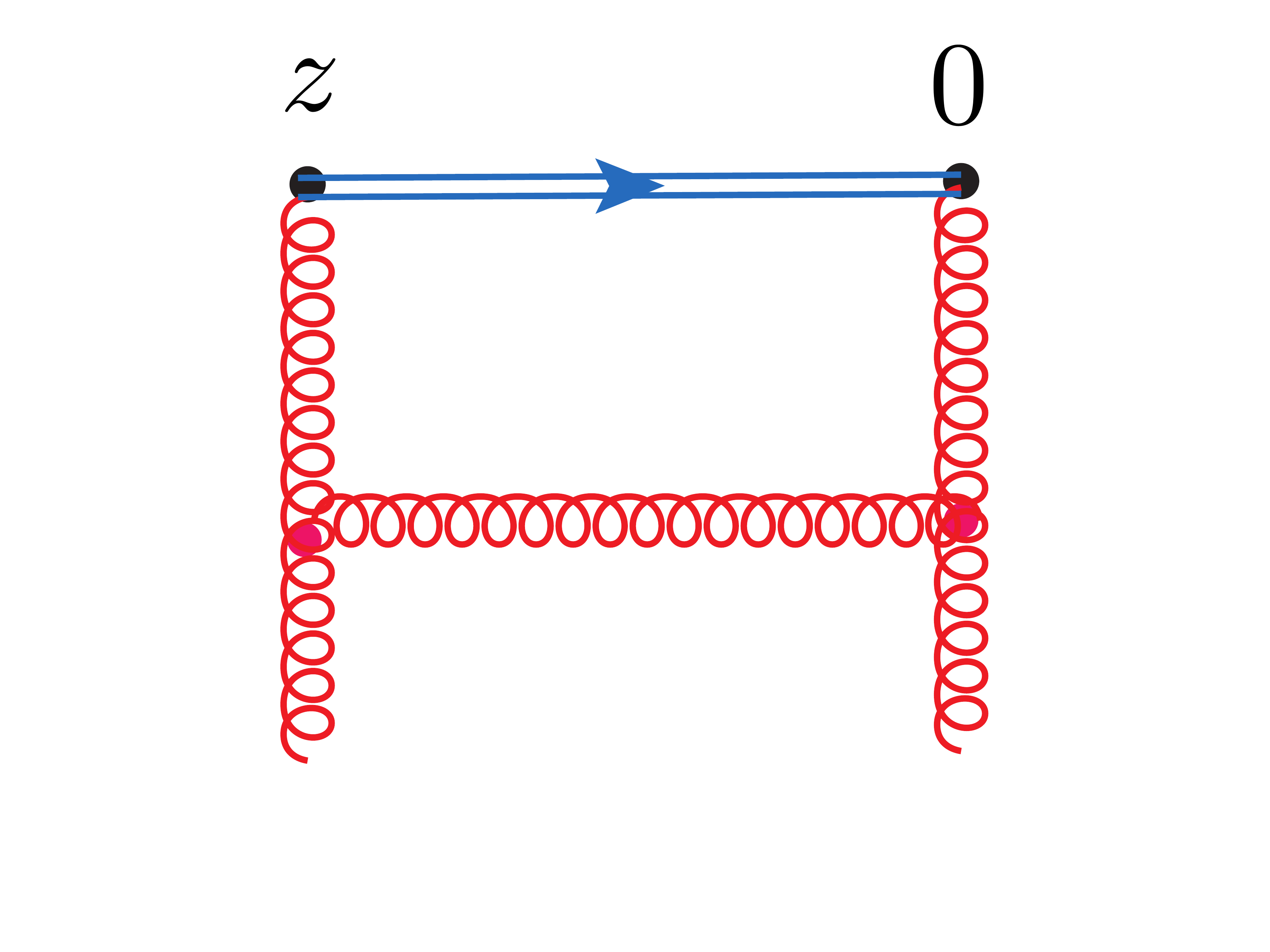}}
        \vspace{-5mm}
   \caption  {Box diagram.     \label{box}}
   \end{figure}

There is also a 
contribution   given by the diagram in Fig.   \ref{box}
containing  a gluon exchange between
two gluon lines.  
This diagram does not have UV divergences, but it has DGLAP %%% -type 
$\ln z_3^2$ contributions.  
In contrast to the vertex diagrams, the original  
$G_{\mu\alpha}( z)  G_{\nu \beta }( 0)$ operator generates 
now a mixture of bilocal operators corresponding 
to various projections of $G_{\mu\alpha}( \bar uz)  G_{\nu \beta }( 0)$
onto the structures built from vectors $p$, $z$ and the metric tensor $g$. 

In particular, in the case of the original 
$\bra{p} G_{0i}( z)G_{0 i}(0)  \ket{p}$
matrix element, the box diagram contribution is expressed through
matrix elements of $\bra{p}G_{0i}( uz)G_{0i}(0) \ket{p} $, $ \bra{p} G_{3i}(uz) G_{3i}(0) \ket{p}   $,
$ \bra{p} G_{30}(uz) G_{30}(0) \ket{p}   $ and 
$\bra{p} G_{ij}(uz) G_{ij} (0)\ket{p} $  types.
All these matrix elements also appear in the box diagram
 if one starts with the 
$\bra{p} G_{3i}( z)G_{3i}(0)  \ket{p}$
matrix element.  Thus,  
 in both cases we have   a complicated 
mixing of different types of operators. 

The situation is simpler for matrix elements 
\begin{align}
M_{03} ^\pm(z,p)  \equiv \bra{p}G_{0i}( z)G_{i3}(0)  \pm G_{3i}(z) G_{i 0}(0) \ket{p}  \ .
\label{M03}
\end{align}
 Namely, for   $M_{03} ^+(z,p) $  (or $M_{03} ^-(z,p) $)  combination,  the box diagram contribution
 is expressed through 
$M_{03} ^+(uz,p) $  (or  $M_{03} ^-(uz,p) $) only. 
  However,   
\begin{align}
M_{03} ^- \equiv 
 { M}_{0 i; i3 } - { M}_{3 i; i 0 } =  & 2 p_0 z_3  \left ( \mathcal{M}_{pz} -\mathcal{M}_{zp} \right)  \  .
\end{align}
 does not  contain the twist-2 function $\mathcal{M}_{pp} $,
 and is of no interest.  
  For $M_{03} ^+(z,p) $,  the box 
contribution   is given by 
\begin{align}
&
  \frac{N_c g^2 \Gamma(d/2-1)}{4\pi^2\left(-z^2\right)^{d/2-2}}   \int_0^1 \dd u  \left
  (\bar uu + \frac{2}{3}   \bar u^3 \right )M_{03} ^+(uz,p)  \nonumber \\ 
  & + \frac{N_c g^2\Gamma(d/2-2)}{4\pi^2\left(-z^2\right)^{d/2-2}} \int_0^1 \dd u   \  [\bar u   (1+u^2) -u ] 
  M_{03} ^+(uz,p)      \ .
  \label{box+}
\end{align}

Here, the $\Gamma(d/2-2)$ terms are singular for $d=4$, which results  in 
$\ln (-z^2)$  terms generating the DGLAP evolution.
The $\Gamma(d/2-1)$  terms are singular for $d=2$, which corresponds 
to the fact that the gluon propagator in two  dimensions 
has a logarithmic $\ln (-z^2)$ behavior in the coordinate space.
For $d=4$, these terms are finite.
Note that, unlike the vertex part, the box  contribution 
does not have the plus-prescription form.  

\setcounter{equation}{0}

\section{Gluon self-energy diagrams}

  \begin{figure}[b]
   \centerline{\includegraphics[width=1.5in]{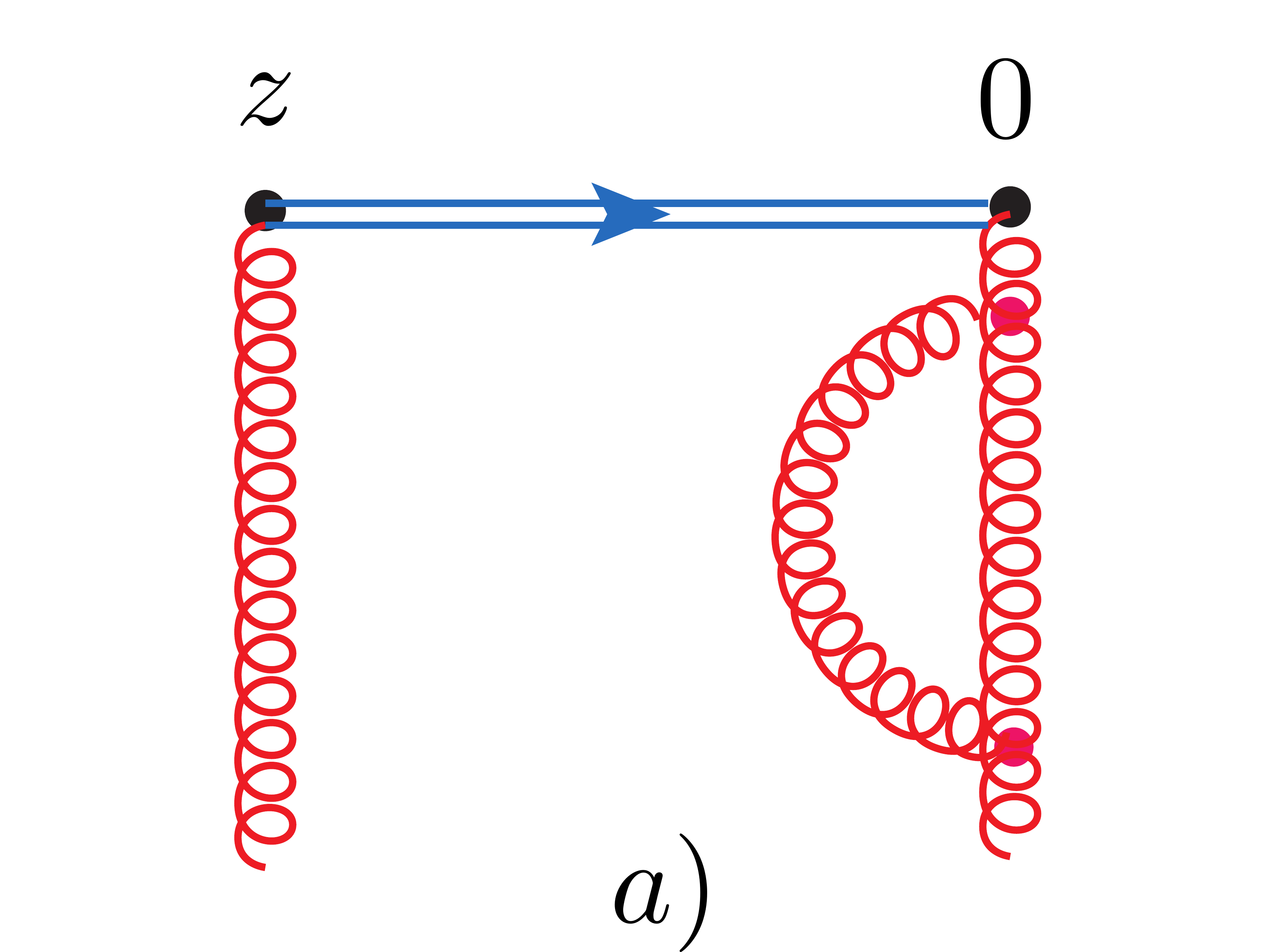} \hspace{-5mm} \includegraphics[width=1.5in]{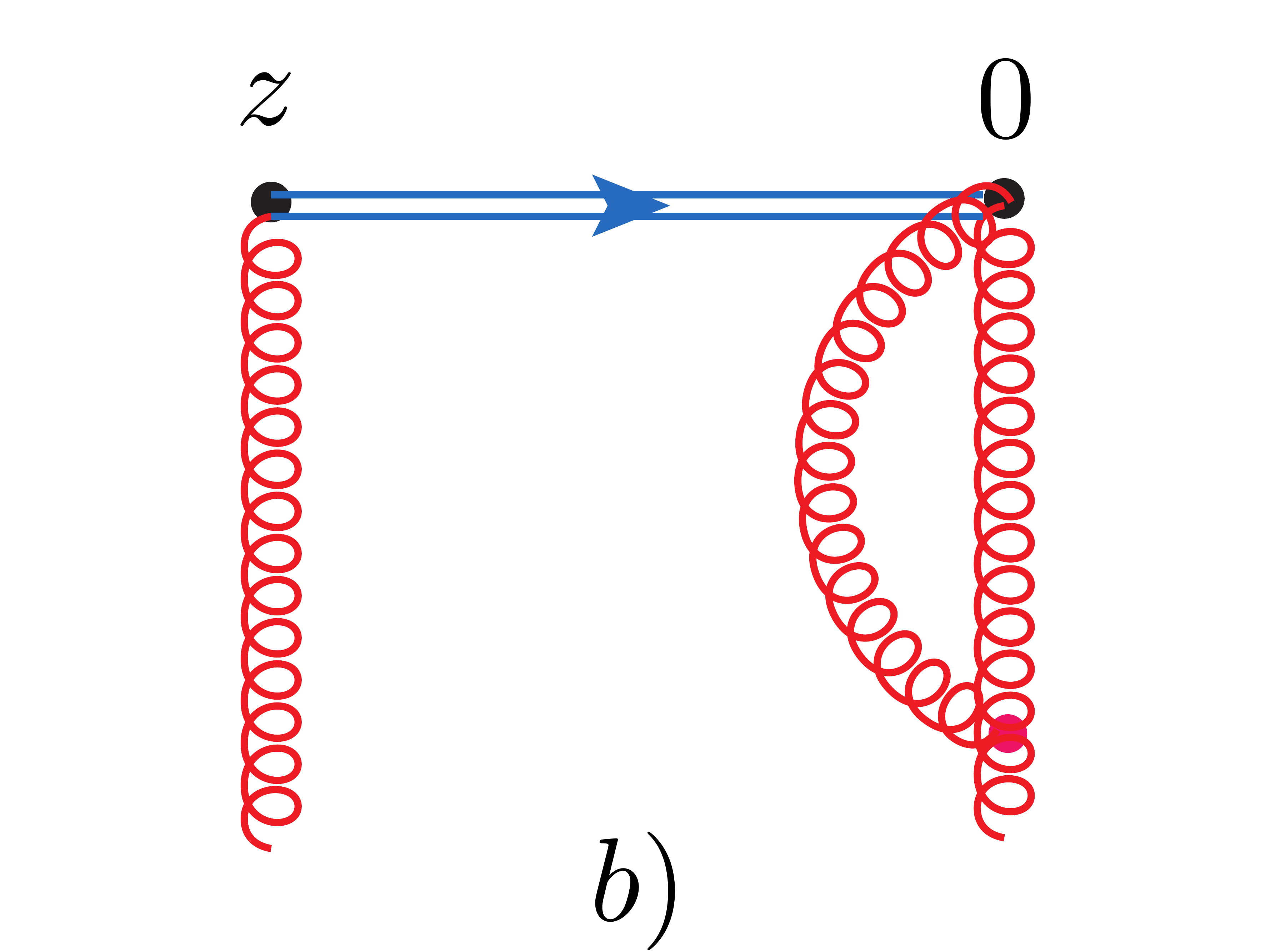}}
   \caption  { Gluon  self-energy-type insertions into the right leg.  
   \label{gluself}}
   \end{figure}

 One may  expect   that
 the plus-prescription form  would appear after the addition of the 
  gluon  self-energy diagrams,  one of which is shown in Fig. \ref{gluself}a.
These diagrams have both the UV and collinear 
divergences. On the lattice, the UV divergence   is  regularized  by  the lattice spacing.
In a continuum theory, one may use the  
 Polyakov prescription  $1/z^2\to 1/(z^2 -a^2)$ for the gluon propagator. 
 The collinear divergences may be regularized by taking 
 a finite gluon mass $\lambda$.  The result is a 
$\ln (a^2 \lambda^2)$  contribution.  
However, it does not have the  $z$-dependence,
and  apparently cannot help one  to build  the  plus-prescription form for the 
 $\ln z_3^2$  part of the box  contribution.

A possible way out is to represent    $\ln (a^2 \lambda^2)$  
 as   the difference   $\ln (z_3^2  \lambda^2)   -  \ln (z_3^2/a^2  ) $ 
of the evolution-type logarithm   $\ln (z_3^2  \lambda^2)   $ 
and a UV-type logarithm  $ \ln (z_3^2/a^2  ) $. 
The latter  can be  added 
to the UV divergences of the diagrams \ref{linkself} and \ref{link}
corresponding to link self-energy and vertex corrections. 
The  $\ln (z_3^2  \lambda^2) $
part  is then  added to the evolution  kernel.

To be on safe side with gauge invariance, we  use  
the dimensional regularization.
Then the analog of the $\ln (a^2 \lambda^2)$ logarithm 
is  a pole $1/(2-d/2)$ sometimes   written as \mbox{$1/\epsilon_{\rm UV} - 1/\epsilon_{\rm IR} $. } 
For our purposes, it is  more convenient to  
 symbolically write it  in a form similar to  $\ln (a^2 \lambda^2)$. 
 Changing $\lambda \to\mu_{\rm IR} $ and $a  \to 1/\mu_{\rm UV}$ we get  $\ln (\mu_{\rm IR}^2/\mu_{\rm UV}^2)$,
and then split this  into the difference $\ln (z_3^2 \mu_{\rm IR}^2)   -  \ln (z_3^2 \mu_{\rm UV}^2 ) $. 

We should also take into account the diagrams (one of them is shown in Fig. \ref{gluself}b) 
with an extra  gluon 
line going out of the link-gluon vertex.  The combined contribution 
of the Fig. \ref{gluself} diagrams and their left-leg analogs  is given by
\begin{align}
{g^2N_c\over 8\pi^2} 
\frac1{2-d/2}
\left [2 - \frac{\beta_0}{2N_c} \right ]G_{\mu\alpha}(z)G_{\lambda \beta }(0) \  ,
\label{counter3}
\end{align}
where $\beta_0 =11N_c/3$ in gluodynamics, so that the terms in the square bracket combine into 1/6.
As discussed above, we will treat   \mbox{$ 1/({2-d/2})$ } as 
\mbox{$\ln (z_3^2 \mu_{\rm IR}^2)   -  \ln (z_3^2 \mu_{\rm UV}^2 ) $. }

\setcounter{equation}{0}

\section{DGLAP evolution structure}

\subsection{ When DGLAP is diagonal in pure gluodynamics 
}

The $M_{03} ^+\equiv 
 { M}_{0 i; i 3 } + { M}_{3 i;i 0 }   $ combination defined by Eq. (\ref{M03}) contains  the twist-2 amplitude $ \mathcal{M}_{pp} $, 
\begin{align}
M_{03} ^+ 
 =   4 p_0     p_3  \mathcal{M}_{pp}   +
 2 p_0 z_3  \left ( \mathcal{M}_{pz} +\mathcal{M}_{zp} \right)  \  ,
\end{align}
though with a higher-twist admixture \mbox{$\mathcal{M}_{zp}+ \mathcal{M}_{pz}$}. 
In the local limit, the relevant operator    is proportional  to the 3rd component of the Poynting vector
\begin{align*}
{\bf S_3}=
({\bf E \times B})_3 =E_1 B_2- B_1 E_2 =-( G_{01} G_{13} +G_{32} G_{20})  \ .
\end{align*}

As already mentioned, the  box part of  the one-loop correction   to the matrix element
$M_{03} ^+(z_3,p)$  
in pure gluodynamics %%%%%%
 has a simple DGLAP structure\footnote{This simplicity may be violated in higher orders.}   (\ref{box+}).
Combining all the  gluon one-loop corrections  to it, 
we get, in the $\overline{\rm MS}$ scheme,
\begin{align}
& \frac{g^2N_c}{8\pi^2}\int_0^1 \dd u  
 \left\{ \left[ \left (\frac{3}{2} -\frac16 \right )  \ln(z_3^2 \mu_{\rm UV}^2 e^{2\gamma_E}/4)  +2 \right] \delta(\bar u)  \right. \nonumber  \\
 &\quad  \left.       -        2  \log(z_3^2  \mu_{\rm IR}^2 e^{2\gamma_E}/4) \left [ \frac{ (1- u \bar u)^2}{\bar u} \right ]_+  \right. 
   \label{M031L} \\
 &\quad \left.      +    \left [ u -3  \frac{u}{ \bar u} -4 \frac{ \log (\bar u)}{\bar u}     \right ]_+ 
  +         2 \left  (\bar uu +  \frac{2}{3}\bar u^3  \right ) \right\}  \ M_{03} ^+(uz,p) \   .\nonumber 
\end{align}

The first line here comes from the UV-singular contributions. It contains the $\delta(\bar u) $ factor
which reflects 
the local nature of the UV divergences and   converts 
$M_{03} ^+(uz,p) $ into $M_{03} ^+(z,p) $. 
The second line contains the Altarelli-Parisi (AP) kernel 
  \begin{align} 
B_{gg} (u)  =&
     2  \left [\frac{(1-u\bar u)^2} {1-u}   \right ]_{+} \ .
     \label{V1}
  \end{align}   
It  has  the plus-prescription 
 structure reflecting the fact that, 
in the local limit,  ${\mathcal{M}}_{pp} (z,p)$  is proportional to   the  
matrix element of the gluon energy-momentum tensor.
From now on,  ``+'' means the plus-prescription at 1.

The third line 
contains $z_3$-independent  terms coming from the vertex diagrams
(these have the  plus-prescription form) and from  the box diagram. 
The latter may be written as a sum of  the term $ 2 \left  (\bar uu +  \frac{2}{3}\bar u^3  \right )_{+} $ 
that has   the plus-prescription form  and the  term $ \frac23 \delta (\bar u) $ that 
may  be combined with the UV terms.

\subsection{Reduced Ioffe-time distribution}

The combination  $ \mathcal{M}_{pz} +\mathcal{M}_{zp}$   is an   odd  function of $\nu =z_3 p_3$.
Writing it as $2 z_3 p_3 m^+_{zp} (\nu, z_3^2)$, we have
\begin{align}
M_{03} ^+(z_3, p) =   4     p_3 p_0 [ \mathcal{M}_{pp}  (\nu, z_3^2 ) +
z_3^2  \   m^+_{zp} (\nu, z_3^2 )] \ .
\label{M03p}
\end{align}
Dividing out the kinematical factor $4     p_3 p_0$, we deal with 
\begin{align}
\widetilde{\mathcal{M}}_{pp}  (\nu, z_3^2 ) \equiv
\mathcal{M}_{pp}  (\nu, z_3^2 ) +
z_3^2  \   m^+_{zp} (\nu, z_3^2 ) \ , 
\end{align}
which is a  function of  $\nu $ and $ z_3^2$.
Now, just like in  the quark case considered in  Refs. \cite{Radyushkin:2017cyf,Orginos:2017kos},  we can introduce 
the reduced Ioffe-time distribution  
 \begin{align}
\widetilde{\mathfrak M} (\nu, z_3^2) \equiv \frac{ \widetilde{\mathcal{M}}_{pp}  (\nu, z_3^2)}{\widetilde{\mathcal{M}}_{pp} (0, z_3^2)} \  . 
 \label{redm0}
\end{align}
Since  $ \widetilde{\mathcal{M}}_{pp}  (\nu, z_3^2)$ is obtained from  the multiplicatively 
renormalizable combination $M_{03} ^+ $, 
the  UV divergent $Z(z_3^2 \mu^2_{UV} )$  factors generated by the link-related  and gluon self-energy diagrams 
cancel in the ratio (\ref{redm0}).  As a result, the small-$z_3^2$ dependence of the reduced 
pseudo-ITD  $\widetilde{\mathfrak M} (\nu, z_3^2) $  comes from the logarithmic DGLAP evolution
effects only. 
Moreover,   $ \widetilde{\mathcal{M}}_{pp}  (0, z_3^2)$  
 has no DGLAP  logarithmic  dependence  on $z_3^2$,
because of the plus-prescription nature of the AP kernel $B_{gg} (u)$. 

 Thus,  neglecting ${\cal O} (z_3^2)$ terms, we conclude that, 
 %%%%%%  {\it in the absence of the quark-gluon mixing,} 
 {\it in pure gluodynamics, } %%%%%%%%%%
 $\widetilde{\mathfrak M} (\nu, z_3^2)  $ satisfies the evolution equation 
    \begin{align}
    \frac{d}{d \ln z_3^2} \,  
\widetilde{\mathfrak M} (\nu, z_3^2)    &= - \frac{\alpha_s}{2\pi} \, N_c
\int_0^1  du \,   B_{gg} ( u )   \widetilde{\mathfrak  M} (u \nu, z_3^2)  
\label{EE}
 \end{align}
with respect to
$z_3^2$.   
This relation is modified when gluon-quark transitions are present. %%%%%%%

%%%%%%   A similar (but more lengthy) equation may be written if  
%%%%%%   the gluon-quark mixing term  is also taken into account.
%%%%%%   We postpone the discussion of this subject to a more detailed publication.

\subsection{Gluon-quark mixing}

\begin{figure}[t]
   \centerline{\includegraphics[width=1.7in]{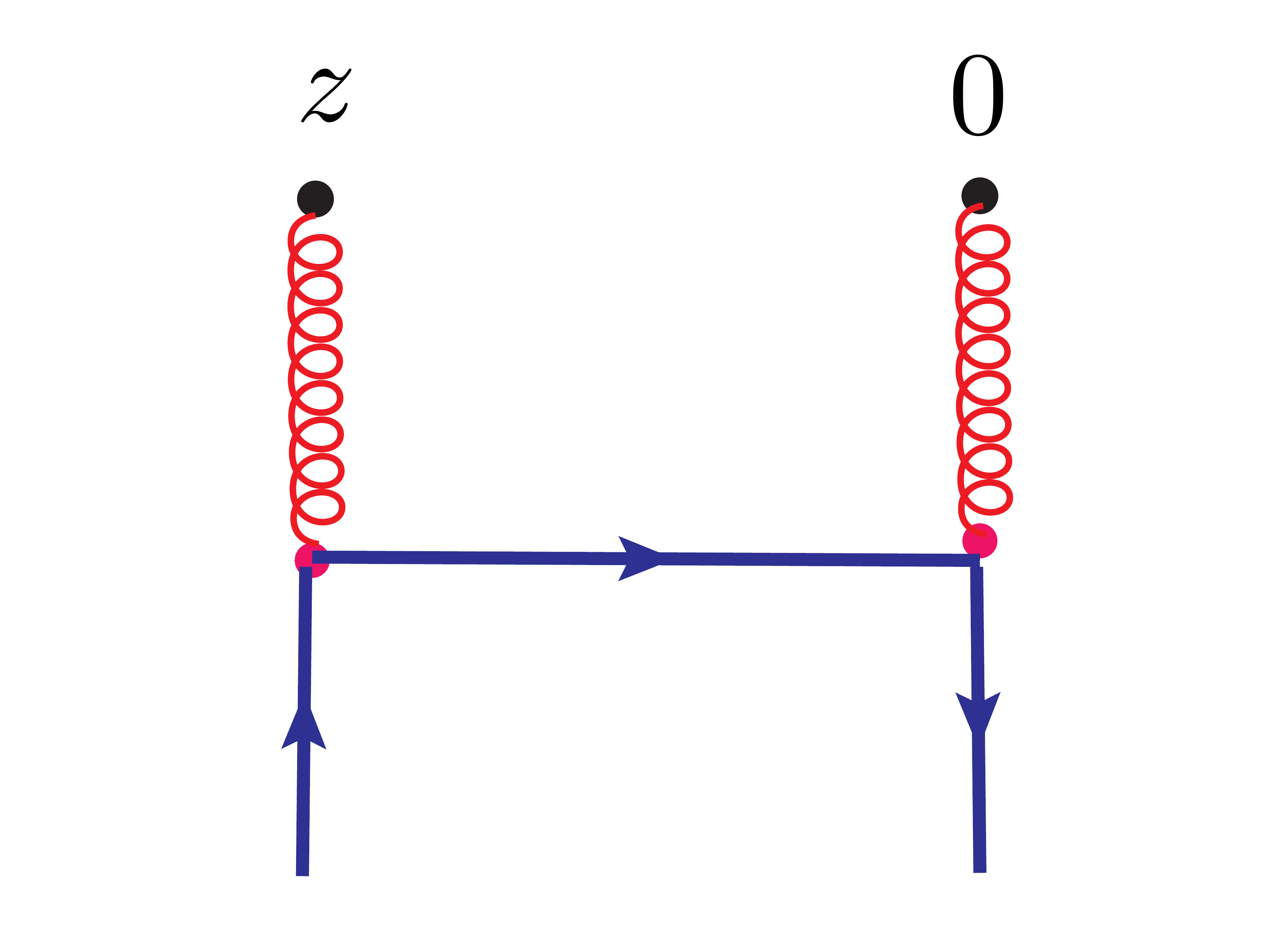}}
        \vspace{-3mm}
   \caption  {Gluon-quark mixing diagram.     \label{gluqum}}
   \end{figure}

In the $\overline{\rm MS}$ scheme, the contribution to $M_{03} ^+$ from  the gluon-quark diagram shown in Fig. \ref{gluqum} 
is given by 
 \begin{align}
&   \frac{ g^2 C_F}{4\pi^2 z_3}   \int_0^1 \dd u  
  \Biggl [ -2u  - \ln(z_3^2 \mu_{\rm IR}^2 e^{2\gamma_E}/4)   \left[2\bar u+\delta(\bar u) \right] \Biggr ] {\cal O}_q (uz_3) \ ,
  \label{gqmix}
\end{align}  
where $ {\cal O}_q (z_3)$  is a  singlet combination of quark fields,
  \begin{align}
& {\cal O}_q (z_3)=  \frac{i}{2} \sum_f  \left(\bar\psi_f( 0)  \gamma_0  \psi_f (z_3) - \bar\psi_f( z_3)  \gamma_0  \psi_f(0)  \right)  \ ,
 \label{gsing}
\end{align}  
with $f$ numerating  quark  flavors. 
Note that  ${\cal O}_q (z_3) $ vanishes 
for $z_3=0$. 
Expanding ${\cal O}_q (z_3) $  in $z_3$
  \begin{align}
& {\cal O}_q ( z_3)=   z_3\,   \frac{i }{2} \sum_f  \bar\psi_f( 0)\gamma_0  \!\stackrel{\leftrightarrow} {\partial}_3\!   \psi_f(0) + {\cal O} (z_3^3)   \ ,
 \label{gsing}
\end{align}  
we see that the lowest term is proportional to the quark part of the energy-momentum tensor.
This term is accompanied  by the $z_3$ factor which cancels the overall $1/z_3$ factor in Eq. (\ref{gqmix}).

The matrix element of ${\cal O}_q ( z_3)$ can be parametrized by 
  \begin{align}
& \langle p| {\cal O}_q ( z_3) |p\rangle =  2  p^0  
\int_0^1 \dd x \, \sin (x p_3 z_3) \, q_S (x)  
 \label{qpar}
\end{align}  
where $f_S (x) = \sum_f [q_f (x) +\bar q_f (x) ]$ is the singlet quark distribution. 
To extract the overall $z_3$  factor, we rewrite 
 \begin{align}
& \langle p| {\cal O}_q ( z_3) |p\rangle
%\nn  & 
= 2  p^0  p_3 z_3
\int_0^1 \dd y \, \int_0^1 \dd \alpha\,  \cos (\alpha y  \nu) \,  y \, f_S (y ) 
 \label{qpar2}
\end{align}  
where $\nu = p_3 z_3$,  as usual.  This gives 
 \begin{align}
&\frac1{z_3}  \int_0^1 \dd u \,  A(u) \, \langle p| {\cal O}_q (u  z_3) |p\rangle
\nn  &
 =  2  p^0  p_3 \,  \int_0^1  \dd w  {\cal I}_S (w \nu)   {\cal A} (w) % \nn & \times 
 \label{qpar3}
\end{align}  
for  $u$-integrals of Eq. (\ref{gqmix}) type. Here 
 \begin{align}
&  {\cal I}_S (\nu)  =
\int_0^1 \dd y \ \cos (y   \nu) \,  y\,    f_S (y) 
 \label{MS}
\end{align}  
is the singlet quark Ioffe-time distribution, and 
 \begin{align}
 {\cal A}  (w)= 
  \int_w^1 \dd u \,  A(u) \ . 
% \nn & \times 
 \label{Aw}
\end{align}  
For the evolution kernel $B_{gq} (u) \equiv   2\bar u+\delta(\bar u)  $, we get
 \begin{align}
 {\cal B}_{gq}  (w)= 
  \int_w^1 \dd u \,  B_{gq} (u) = 1+ (1- w)^2 \  .
% \nn & \times 
 \label{Bw}
\end{align}  

\subsection{Matching relations}

A disadvantage of the $M_{03} ^+(z_3, p)$ combination  is that 
%%%%%%%.   for kinematic reasons, 
 it vanishes 
 when  $p_3=0$ (see \mbox{Eq. (\ref{M03p})).}
Thus, to extract 
$\widetilde{\mathcal{M}}_{pp}  (\nu, z_3^2 )$   
 for $\nu=0$, 
one should
make measurements of  $M_{03} ^+(z_3, p)$ for a few low values of  $p_3$, 
%%% and 
divide $p_3$ out,  %%%%% of the results,
and extrapolate the results to \mbox{$p_3=0$.}  This procedure leads %%%%%%
to additional systematic uncertainties.  %%%%%%%%%

   %\vspace{-5mm} 

Fortunately,   %%%%%
the combination 
$
 { M}_{0 i;  i 0 }  -  { M}_{ i j; i j} =   2   p_0^2  \mathcal{M}_{pp}   $ of \mbox{Eq. (\ref{00m}),} being proportional 
 to $p_0^2$,  does not  have this problem.
 Furthermore, it  gives the twist-2 amplitude $ \mathcal{M}_{pp}$ without contaminations.
 The amplitude  $ \mathcal{M}_{pp} (\nu, z_3^2)$ obtained in this way
 may be used 
  to form the reduced pseudo-ITD
 ${\mathfrak M} (\nu, z_3^2) $, as in Eq. (\ref{redm0}).
 
 %%%%%  Another advantage of this choice %%%%%%%%
 %%%%%  comes from the fact  that the gluon-quark mixing term, due to Eq.(\ref{qpar2}),  is proportional  %%%%%
%%%%%%%  to $p_3$. Hence, it   vanishes  %%%%%
%%%%%   for $p_3=0$, thus making no contribution to the denominator factor  %%%%%
 %%%%%   built for $p_3=0$.  %%%%%

  Using the   results of our  calculations for the one-loop corrections to
 $
 { M}_{0 i; i 0 }  $ and $  { M}_{ i j; i j}$,  and keeping just the $
 { \cal M}_{pp}$ term in the correction (while skipping the ``higher  twist'' terms
$ {\cal M}_{zz},  {\cal M}_{zp},  {\cal M}_{pz},  {\cal M}_{ppzz}$) 
  we obtain 
 the matching  relation 
\begin{align}
 {\mathfrak M}& (\nu, z_3^2)\,  {{\cal I}_g (0, \mu^2) } = { {\cal I}_g (\nu, \mu^2) }
-   \frac{\alpha_s N_c}{2\pi}\int_0^1 \dd u \,  {{\cal I}_g (u \nu, \mu^2)}
 \nonumber  \\
 & \times 
 \Biggl \{     \ln (z_3^2  \mu^2 e^{2\gamma_E}/4)  \, B_{gg} (u) +4 \left [ \frac{u+ \log (\bar u)}{\bar u}\right ]_{+}  \nonumber \\
 &   +  \frac23   \left [ 1 -u^3
     \right ]_+  \Biggr \}  
   -   \frac{\alpha_s C_F}{2\pi}   \ln (z_3^2  \mu^2 e^{2\gamma_E}/4) 
   \nonumber  \\
 & %%%%  \hspace{5mm} 
   \times 
 %%%%%     \frac{p_3}{p_0}
   \int_0^1 \dd w \, 
    \left [{\cal I}_S (w \nu, \mu^2) - {\cal I}_S (0, \mu^2) \right ]%%%%%%
 \, {\cal B}_{gq} (w) \ 
  \label{matching} 
\end{align}
between   the ``lattice function''   
 ${\mathfrak  M}(  \nu,z_3^2) $ and the light-cone ITDs  $  {\cal I}_g (\nu,\mu^2)$  and 
 $  {\cal I}_S(\nu,\mu^2)$. The first of them is 
 related to the  gluon PDF  ${f}_g(x,\mu^2)$
by 
   \begin{align}
  {\cal I}_g (\nu,\mu^2) =\frac12 
   \int_{-1}^1 dx \,  \, 
e^{ix\nu} \,x  {f}_g(x,\mu^2)  \   .  
 \label{If}
\end{align}
Since $x  {f}_g(x,\mu^2) $ is an even function of $x$,
the real part of $  {\cal I}_g (\nu,\mu^2)$ is given by the cosine transform of $x  {f}_g(x,\mu^2) $,
while its imaginary part vanishes. 
The factor ${\cal I}_g (0, \mu^2) $  has the meaning of the  fraction of the hadron 
 momentum carried by the gluons, ${\cal I}_g (0, \mu^2)  =  \langle x \rangle_{\mu^2}$. \
 %   \begin{align}
 %{\cal I}_g (0, \mu^2) = \int_0^1\dd x \,  x f_g (x,\mu^2) \equiv   \langle x \rangle_{\mu^2} \ .
 %\end{align} 
 
 Thus, Eq. (\ref{matching})  allows to extract the shape of  the gluon distribution.
 Its normalization, i.e., the value of $ \langle x \rangle_{\mu^2} $  should be 
 found by an independent  lattice calculation, similar to that performed 
 in Ref. \cite{Yang:2018bft}.
 The singlet quark function ${\cal I}_S (w \nu, \mu^2)$ that appears in the  %%%%%%%
 ${\cal O} (\alpha_s)$ correction should be also calculated (or estimated)     %%%%%%%%%%
 independently.  %%%%%%%%%%

 Substituting Eq. (\ref{If}) into the matching condition (\ref{matching}), we  can 
 rewrite  the latter in the kernel form \cite{Radyushkin:2019owq}
     \begin{align} 
 {\mathfrak M}(\nu,z_3^2)     &  = 
  \int_{0}^1 \dd x \, 
\frac{x {f} _g (x,\mu^2) }{ \langle x \rangle_{\mu^2}}   \, R_{gg} (x \nu, z_3^2 \mu^2 ) \nn & 
+   %%%%%   \frac{p_3}{p_0} 
 \int_{0}^1 \dd x \, 
\frac{x {f} _S (x,\mu^2) }{ \langle x \rangle_{\mu^2}}   \, R_{gq} (x \nu, z_3^2 \mu^2 )  \ , 
 \  
\label{MtchI}
 \end{align}
 where the kernel $R_{gg} (x \nu, z_3^2 \mu^2 )$ 
  is given by
    \begin{align} 
 R_{gg}(y, z_3^2 \mu^2)   =& \cos y  
  - \frac{\alpha_s}{2\pi} \, N_c 
 \Biggl \{    \ln \left ( z_3^2 \mu^2\frac{ e^{2\gamma_E+2}}{4}  \right )  \,
  R_B(y)\nn & +R_U (y) 
+R_L(y) +R_C(y)  \Biggr \} \,  ,
\label{MtchR}
 \end{align}
 with $R_B(y)$ being 
 the cosine Fourier transform of the $ B_{gg}$  kernel 
 \begin{align}
R_B(y) =& \int_0^1 \dd u \,   B_{gg} (u) \, \cos (u y) \ .
  \label{RB} 
\end{align}
Its  calculation is straightforward, and the result is expressed in terms of  $\cos y$, $\sin y$
and the  integral cosine $\text{Ci}(y)$  and sine  $\text{Si}(y)$  functions. 
The latter come from the $1/(1-u)$ part  of $B(u)$, which gives
 \begin{align}
 \int_0^1 \dd u \, &  \left[ \frac1{1-u} \right ]_+  \,  \cos (u y) = \sin (y) \,  \text{Si}(y)  \nn & +\cos (y)\,
  [\text{Ci}(y)-\log (y)-\gamma_E ]
  \ .
  \label{RB0} 
\end{align}
This combination also appears in $R_U(y)$, which is given by the cosine transform of $4[u/\bar u]_+$.  
Similarly,  $R_L(y) $ is the cosine transform of the \mbox{$4[(\ln \bar u)/\bar u]_+$} term.
It involves a hypergeometric function 
   \begin{align} 
 R_L(y) = 
 4\,  {\rm Re} \left [ i y e^{i y} \,
   _3F_3(1,1,1;2,2,2;-i y) \right ] \, . 
\label{MtchRL}
 \end{align}
 
The  $R_C(y)$ and $R_{gq} (y)$  kernels are given by the cosine transforms of  $ \frac23   \left [ 1-u^3
     \right ]_+ $ and $\left [ 1+\bar u^2\right]_+$, respectively. 
Expressions for them 
involve  $\cos y$, $\sin y$ and inverse powers of $y$. 

The  important  point is that  the $R(y, z_3^2 \mu^2)$ kernels  are given by {\it explicit}   perturbatively calculable 
expressions. Using them  and Eq. (\ref{MtchI}) one may  directly  relate 
 ${\mathfrak M}(\nu,z_3^2) $  and the light-cone 
PDFs. % $f_g(x, \mu^2)$.  
 Then, 
 assuming some parameterizations  for the $f_g(x,\mu^2)$ and $f_S(x, \mu^2)$ distributions,
 one can  fit  their    parameters and $\alpha_s$ 
 from   the  lattice data for ${\mathfrak M}(\nu,z_3^2)$ using  Eqs. (\ref{MtchI}), (\ref{MtchR}). This procedure is essentially the same as that used 
in  the  ``good lattice cross sections'' approach  \cite{Ma:2014jla,Ma:2017pxb}. 
 
 \subsection{Matching relations for quasi-PDFs}
 
 The kernel relation  (\ref{MtchI})  directly connects ${\mathfrak M}(\nu,z_3^2) $  
 %%% and $f_g(x, \mu^2)$. 
 with PDFs.   %%%%%%
 So, there is no need to introduce 
 intermediate functions, such as quasi-PDFs.
Still,  our results for particular matrix elements,
such as Eq. (\ref{M031L}) for $M_{03} ^+(z_3,p)$,
may be used to get matching conditions for quasi-PDFs. The latter 
are generically defined  \cite{Ji:2013dva} as
     \begin{align} 
Q(y, p_3)      = \frac{p_3}{2 \pi} 
  \int_{-\infty}^\infty \dd z_3  
{\cal M} (z_3, p)   \, e^{-i y p_3 z_3} \ .
 \  
\label{qPDF}
 \end{align}
 
To proceed, one should write the amplitudes  ${\cal M} (z_3, p)  $ through the kernel
relation (\ref{MtchI}) with  $R(x \nu, z_3^2 \mu^2 )$  expressed    in terms of $p_3$ and $z_3$,
call it  $J(x, p_3, z_3)$.
The structure of its dependence on $z_3$ at one loop may be read off Eq. (\ref{M031L}).
For the $gg$ part, 
    \begin{align} 
   &  J_1^{gg} (x, p_3, z_3)= \left (\gamma_U \ln z_3^2   +C_U \right) e^{i  x  p_3 z_3} 
    \nn & - \int_0^1 \dd u \,\left [  \ln z_3^2  \, B_{gg} (u)
    +C (u) \right ]e^{i u x  p_3 z_3} \  .
\label{Jk}
 \end{align}
 The 1-loop quasi-PDF matching kernel is then given by 
  \begin{align} 
Z_1^{gg} (y, x, p_3)      = \frac{p_3}{2 \pi} 
  \int_{-\infty}^\infty \dd z_3  \,  J_1^{gg} (x, p_3, z_3) 
\, e^{-i y p_3 z_3} \  . 
 \  
\label{qZ}
 \end{align}
 The $C_U$ and $C(u)$  contributions  of $  J_1^{gg} (x, p_3, z_3)$ 
 produce \mbox{$C_U \delta (y- x)$} and $C (u)\delta (y - u x ) $ terms.
 Hence, 
 the  resulting parts of   $Q(y,p_3)$ are visible 
 in the ``canonical'' $0\leq |y| \leq 1$ region only.   However, the terms
 with $\ln z_3^2$ give nonzero contributions in the $y/x>1$ 
 and  $y/x<0$ regions %%%%%%
%%%%   region
  as well, namely
    \begin{align} 
   &  \, Z_1^{gg}  (y, x, p_3) |_{y/x>1}  = \frac1{|x|} \left [- \frac{\gamma_U}{\eta -1}    
     + \int_0^1 \dd u   \, \frac{B_{gg} (u)}{\eta-u} \right ]
    \  ,
\label{Zy}
 \end{align} 
 where $\eta= y/x$. 
 The  $Z_1^{gg}  (y, x, p_3) |_{y/x<0}$ term is given by the same expression, but with the opposite sign. %%%%%%
 Note that these contributions are completely
 determined by  the AP kernel $B_{gg} (u)$ and the UV constant 
 $\gamma_U$.  
 Knowing them, one derives from Eq. (\ref{Zy})  a general constraint on the results 
for $Z_1^{gg} (\eta, 1, p_3) |_{\eta>1}$  
and $Z_1^{gg} (\eta, 1, p_3) |_{\eta<0}$     %%%%%%
obtained by any  Feynman diagram calculation. 
 Using explicit form of $B_{gg} (u) $, we find 
  \begin{align}     
    \int_0^1 \dd u \, \frac{ B_{gg} (u) }{\eta-u}   & = 2 \frac{(1-\eta \bar \eta)^2} {\eta -1}  \,  \ln \frac{\eta-1}{\eta}
    \nn &  + \frac{11}{6} \frac{1}{\eta -1} +  \eta (2 \eta -1) +\frac{11}{3}
    \  . 
\label{ZB}
 \end{align} 
 
 For large $\eta$, this expression tends to zero  as ${\cal O}(1/\eta^2)$.
 It should  be stressed that such a behavior results   from   any kernel $B(u)$ that has the plus-prescription form.

 This observation and the explicit expression   given by  \mbox{Eq. (\ref {ZB}) } 
  may be used to check the gluon-gluon matching kernels 
 in \mbox{Refs. \cite{Wang:2017qyg,Wang:2019tgg}. }
 Our check shows that  $ Z_1^{gg}  (y, 1, p_3) |_{y>1}$ corresponding  to Eq. (64) of  Ref. 
  \cite{Wang:2019tgg} does not satisfy  the  constraint (\ref{Zy}).  
 The  difference is  by a constant term (-2/3) that  leads to a linear divergence  
  in the integral of   $ Z_1^{gg}  (y, 1, p_3) |_{y>1}$  \mbox{over $y$. }
  The same difference (with the opposite sign) appears %%%%%%%
  in the $ Z_1^{gg}  (y, 1, p_3) |_{y<0}$ term in Eq. (64) of Ref.  \cite{Wang:2019tgg}. %%%%%%%%%%
Apparently, these differences      result  from   the use of  %%%%%%%%%%
  the  off-shell  external gluon fields  in the calculations of Ref.   \cite{Wang:2019tgg}, %%%%%%
   but  a discussion of  this  topic is outside of   the scope of our paper.  %%%%%%%%%%
 
\section{ Summary.}

In  this paper, we have presented the results that form the basis for 
the ongoing  efforts to calculate gluon PDF using the pseudo-PDF approach.

In particular, we gave a classification of possible two-gluon correlator functions.
We have identified those of them  that  contain  the invariant amplitude ${\cal M}_{pp}(\nu,-z^2)$ 
 that  determines  the gluon PDF  in the light-cone $z^2 \to 0$ limit. 
Since this limit is singular, one needs the matching conditions
that relate  ${\cal M}_{pp}(\nu,z_3^2)$ to the light-cone PDF $f(x,\mu^2)$.

To this end, using  the method  of Ref. \cite{Balitsky:1987bk},  we
have  performed calculations of  the one-loop corrections 
to the  gauge-invariant correlator of two gluon field-strength tensors,
with all Lorentz indices explicit.
To preserve gauge invariance, we have used the dimensional regularization. 

Since the DR produces the same form $\ln z_3^2 \mu^2$ both  for 
logarithms related to the UV singularities 
and for those reflecting the DGLAP evolution, we have made an effort to
separate  these two sources of the \mbox{$\ln z_3^2$-dependence}
at small   $z_3^2$.  
When we form a reduced ITD $ {\mathfrak M}(\nu,z_3^2) $, the UV-related contributions are canceled, and 
only the DGLAP-related terms remain in the matching relation 
between the reduced ITD and the light-cone ITD. 

The matching relation  may be also written in a kernel form (\ref{MtchI})
that directly connects lattice data on $ {\mathfrak M}(\nu,z_3^2) $
with the  normalized 
gluon PDF $x f_g (x,\mu^2)/ { \langle x \rangle_{\mu^2}}  $.
The average gluon momentum fraction $ { \langle x \rangle_{\mu^2}}  $
needs to be extracted from a separate lattice calculation.

We have also demonstrated that our results 
may be used for a   rather   straightforward calculation  
of the 
one-loop corrections to  quasi-PDFs, 
providing new insights concerning   their structure
that may be used to check the results 
for the gluon quasi-PDF matching conditions.

In a future  publication, we plan to present more details 
of our calculations,   and to give a complete result for the box diagram,
in particular for the non-forward kinematics
that are needed in lattice calculations
of distribution amplitudes and GPDs.
We also plan to include  calculations for gluon-quark 
and quark-gluon terms.

{\bf Acknowledgements.} We  thank K. Orginos,  \mbox{J.-W. Qiu}, D. Richards    and S. Zhao for 
interest   in  our   work and  discussions. This work is supported by Jefferson Science Associates,
 LLC under  U.S. DOE Contract \#DE-AC05-06OR23177
 and by U.S. DOE Grant \#DE-FG02-97ER41028. 

%\end{document} 

	\end{document}